\documentclass[3p,times,preprint]{elsarticle}

\usepackage[UKenglish]{babel}
\usepackage[colorlinks=true,urlcolor=blue,citecolor=blue,linkcolor=blue]{hyperref}
\biboptions{sort&compress}

\usepackage{varioref,amssymb,textcomp,bm,amsfonts}
\usepackage{amsmath}
\usepackage{graphicx}
\usepackage{keyval, everysel, ragged2e}

\newcommand{\fig}[1]{fig. \ref{fig:#1}}

\begin{document}

\begin{frontmatter}

\title{Optimised simulated annealing for Ising spin glasses}

\author{S.V. Isakov \fnref{fn1}}
\author{I.N. Zintchenko}
\author{T.F. R\o nnow}
\author{M. Troyer}

\fntext[fn1]{Current address: Google, Brandschenkestrasse 110, 8002 Zurich, Switzerland}

\address{Theoretische Physik, ETH Zurich, 8093 Zurich,  Switzerland}

\begin{abstract}
We present several efficient implementations of the simulated
annealing algorithm for Ising spin glasses on sparse graphs. In
particular, we provide a generic code for any choice of couplings, an
optimised code for bipartite graphs, and highly optimised
implementations using multi-spin coding for graphs with small maximum
degree and discrete couplings with a finite range. The latter codes
achieve up to 50 spin flips per nanosecond on modern Intel CPUs. We
also compare the performance of the codes to that of the special
purpose D-Wave devices built for solving such Ising spin glass
problems.
\end{abstract}


\end{frontmatter}

\noindent
{\bf PROGRAM SUMMARY}\\
\begin{small}
\noindent
{\em Program Title:} SimAn v1.0                               \\
{\em Journal Reference:}                                      \\
{\em Catalogue identifier:}                                   \\
{\em Licensing provisions:} GPLv3                             \\
{\em Programming language:} C++, OpenMP for parallelization.  \\
{\em Computer:} any PC.                                       \\
{\em Operating system:} Linux/OS X/UNIX.                      \\
{\em Has the code been vectorized or parallelized?:} parallelized using OpenMP. \\
{\em RAM:} Variable, from a few megabytes.                    \\
{\em Number of processors used:} Variable.                    
{\em Classification:} 4.13, 6.5, 23.                          \\
{\em Nature of problem:} Ising spin glass ground states on sparse graphs. \\
{\em Solution method:} Simulated annealing.                   \\
{\em Running time:} From milliseconds to seconds.             \\
\end{small}
 
\section{Introduction}

First introduced three decades ago~\cite{Kirkpatrick13051983},
simulated annealing is a powerful algorithm commonly used for
heuristic optimisation due to its simplicity and effectiveness. Within
this approach, variables to be optimised are viewed as the degrees of
freedom of a physical system and the cost function of the optimisation
problem as the energy. One then performs a Monte Carlo simulation of
that system, starting at high temperatures and slowly lowering the
temperature during the simulation, so that ultimately the
configuration of the system ends up in a local minimum. Annealing slow
enough and with multiple repetitions, one can hope to find the global
minimum.

In this paper we present highly optimised implementations of simulated
annealing for the Ising spin glass problem
\begin{equation}
  H = \sum_{i < j} J_{ij} s_i s_j + \sum_{i} h_i s_i
  \label{eq:Hamiltonian}
\end{equation}
where $s_i = \pm 1$. The couplings $J_{ij}$ induce a graph structure
with the spins represented as vertices and with edges between all
neighbour pairs $i$ and $j$ for which $J_{ij} \ne 0$.

The broad interest in the Ising spin glass comes from the fact that
finding the ground state is non-deterministic polynomial (NP) hard
~\cite{Barahona1982}. This means that many other interesting problems,
including constraint satisfaction and the travelling salesman problem,
can be mapped to such Ising spin glasses in polynomial time.

The complexity of solving this problem is also the motivation behind
special purpose devices built by the company D-Wave systems, which can
to date solve Ising spin glass problems with up to $N=512$ spins on
the so-called ``chimera graph'' (discussed in \ref{sec:chimera}). The
existence of these devices has triggered increased efforts into
effectively mapping non-linear combinatorial optimisation problems
from application domains, such as image recognition, to Ising spin
glass problems \cite{0804.4457,1205.1148}. A recent study
\cite{McGeoch} has compared the performance of a D-Wave device against
three general purpose classical optimisation algorithms and concluded
that the D-Wave device tested was 3600 times faster than the fastest
of these codes. In their conclusions, the authors qualify their result
with the statement ``It would of course be interesting to see if
highly tuned implementations of simulated annealing could
compete''. The optimised simulated annealing codes for Ising spin
glasses presented in this publication can also be applied to the
chimera graph of the D-Wave devices and provide such competitive
highly tuned implementations.

Another state-of-the-art approach for finding ground states of spin
glasses is parallel tempering \cite{tempering, Geyer}. This method
can be more efficient than simulated annealing in some cases
\cite{Roma20092821,doi:10.1142/S0129183103004498}, e.g.
for Ising spin glasses with Gaussian distribution of couplings.
Most of the optimisation techniques presented in this paper can also
be applied to parallel tempering.

\section{Optimisations}

Simulated annealing is simple and can be implemented in a short time
for the Ising spin glass. However, a range of optimisations can
improve its performance by orders of magnitude. In this work we
discuss many of these optimisations and present efficient
implementations for modern CPUs in a freely available software
package.

\subsection{Forward computation of $\Delta E$}

Performing simulated annealing using the Metropolis algorithm requires
calculating the acceptance ratio $\exp(-\beta \Delta E_i)$, where
$\beta$ is the inverse temperature and $\Delta E_i$ the energy change
upon flipping the $i$'th spin. The value of $\Delta E_i$ is typically
computed by traversing the neighbours of spin $i$ and takes up most of
the time required for each spin update. However, as for typical
annealing schedules the average acceptance rate is only around $15\%$,
it is much more efficient to calculate and store $\Delta E_i$ for
every spin and only update this value if spin $i$, or one of its
neighbours, is flipped. This way the number of operations if a
spin-flip is accepted is the same with an additional array access. On
the other hand, if a flip is not accepted, $\Delta E_i$ does not have
to be computed, but simply retrieved from an array.

\subsection{Fixed loop lengths and unrolling}
\label{sec:fixed:loops}

One has to loop over neighbours to compute the energy change $\Delta
E_i$ when flipping the $i$-th spin. This loop can be optimised using
fixed loop lengths by specifying the maximum number of neighbours at
compile time. In this case, the compiler can unroll the loop more
efficiently. This approach is advantageous when the distribution of
the number of neighbours is narrow. For instance, for perfect chimera
graphs with five and six neighbours (there might be a few sites with
four neighbours in depleted graphs) the code with fixed loop length is
20\% faster. However, using the fixed loop length codes might be
disadvantageous when the distribution of the number of neighbours is
wide, say, for graphs with the majority of sites having three
neighbours and a few sites having ten neighbours.

\subsection{Fast random number generators}

For a simple model, like the Ising model, generation of random numbers
can take up a substantial fraction of the computational effort. Unlike
simulations aiming at high accuracy results for physical properties,
in optimisation algorithms such as simulated annealing the quality of
the random number generator is not very critical and thus fast
generators, such as the linear-congruential, can be used.

\subsection{Deterministic traversal over lattice sites}

Lattice sites can be picked up for an update sequentially in some
specified order instead of picking them up in random order. This
approach decreases the amount of generated random numbers and leads to
faster codes. Even though the detailed balance condition is violated,
it typically yields betters success rates.

\subsection{Precomputing random numbers}

Random numbers can also be reused across multiple repetitions of the
annealing, as long as they start from different initial
configurations. Furthermore, we can modify the Metropolis acceptance
criterion from $\exp(-\beta \Delta E_i) < u$, where $u\in[0,1)$ is
  drawn uniformly at random, to a cheaper decision $\Delta E < r$,
  where $r=-\frac{1}{\beta}\log{u}$. The values $r$ can then be stored
  instead of $u$. For the case of integer couplings we can further
  optimise by using integer comparison instead of floating point
  comparisons.

Correlations introduced by reusing random numbers can be significantly
reduced with minimal additional effort by cyclically shifting the
precomputed array of random numbers for each sweep (a sweep is defined
as one attempted update per spin) by a random offset. We observed that
the remaining correlations have only a minimal impact on the
performance of the simulated annealer.

\subsection{Optimisations for bipartite graphs}

If the graph is bipartite, the complexity of finding the ground state
configuration can be reduced to finding the optimal spin-vector for
only one sub-lattice. Lets split the set of spins into two sets $A$
and $B$ such that the spins from one set only couple to spins of the
other. Without loss of generality we assume that all on-site fields
$h_i = 0$, $N_A \leq N_B$ and lets sort the spins such that all spins
in $A$ come before those in $B$. We use the notation $\mathbf{s}_A =
\{s_1,s_2,\ldots,s_{N_A}\}^T$, $\mathbf{s}_B =
\{s_{N_A+1},s_{N_A+2},\ldots,s_{N}\}^T$ and $\mathbf{s} =
\{\mathbf{s}_A,\mathbf{s}_B\}$. The couplings $J_{ij}$ are then in
matrix form
\begin{equation}
J = 
\begin{pmatrix}
\mathbf{0} & C^T \\
C & \mathbf{0}
\end{pmatrix}
\end{equation}
where the energy can be calculated as
\begin{equation}
  E = \frac{1}{2} \mathbf{s}^T J \mathbf{s} = \frac{1}{2}\left( \mathbf{s}_A^T  C^T  \mathbf{s}_B + \mathbf{s}_B^T C \mathbf{s}_A \right) = \mathbf{s}_B^T  C  \mathbf{s}_A.
\end{equation}
which can be minimised by finding the optimum $\mathbf{s}_A$ such that
\begin{equation}
  E = \min_{\mathbf{s}_A} \left\{ -\sum_{i=0}^{N_B}\left|\sum_{j=0}^{N_A}C_{ij} s_j\right| \right\}
\end{equation}
as we can always align the spins in $\mathbf{s}_B$
accordingly. Therefore, for bipartite graphs we need to simulate and
update only $N_A \le N/2$ spins. In the above argument we for
simplicity made the assumption that all on-site fields are zero. If
this is not the case, we can transform the Hamiltonian by introducing
two ancillary spins, one coupled to all spins in sub-lattice $A$ with
couplings $h_i$ for $i \in [1,N_A]$ and the other to all spins in
sub-lattice $B$ with couplings $h_i$ for $i \in [N_A+1,N]$. The two
spins are also coupled to each other with a strong ferromagnetic
bond. After this transformation, the graph remains bipartite, but all
on-site fields are expressed as couplings between spins and the above
argument can be applied.

\subsection{Multi-spin coding}
\label{sec:multi-spin-coded}
In contrast to the standard implementations of simulated annealing where
one uses an integer to store every spin, higher efficiency can be
archived by representing spins by 
single bits which allows one to update many spins
simultaneously. This approach is known as {\it multi-spin coding}. We
here present two different implementations of multi-spin coded
simulated annealers. The codes were written for different ranges of
couplings and with up to six nearest neighbours, using words of $S=64$
bits to stores $64$ spins. Rather than storing spins from a single
lattice across a word, as described in \cite{Block20101549}, we store
$64$ replicas of the same spin in one word.

\subsubsection{Approach one}
\label{sec:approach-one}
Multi-spin coded versions of simulated annealing were first suggested
in Ref.~\cite{Bhanot} and later extended in Refs.~\cite{130097,
  Rieger1993,msc_ref}. The implementation presented here is based on
the outline found in Ref.~\cite{msc_ref}. For completeness, we
summarise this approach. In a standard Monte Carlo simulation of the
Ising model, one selects a random spin and computes the energy $\Delta
E$ gained by flipping this spin. The move is either accepted or
rejected with a probability given by the Boltzmann factor $e^{-\beta
  \Delta E}$, where $\beta$ is the inverse temperature. In the
approach suggested in Ref.~\cite{msc_ref}, this part of the algorithm
is slightly altered.  For a finite number of lattice neighbours and
integer couplings, the number of all possible values of $\Delta E$ is
finite and these values can be ordered in descending order $\Delta E
\in \{-2E_0,\dots, -2E_m \}$, where $E_i$ are the local energy levels
of the spin and $m+1$ is the number of energy levels.
Instead of calculating $\Delta E$, one can efficiently
calculate the index $i$ of $-2E_i$ for all the $64$ spins
simultaneously by bitwise summation, i.e. indices are stored in
$\lceil \log_2(k) \rceil$ words, where $\lceil\, \rceil$ denotes the
next largest integer value. The spins which need to be flipped with
probabilities $p_{t,i} = e^{2\beta E_{i}}$ can be determined by
performing simple boolean logic on words that represent indices $i$
and by comparing the probabilities $p_{t,i}$ with a uniformly
distributed random number $0\leq u < 1$ starting at the highest through
the lowest level. This can be illustrated by the following pseudo code
example for a spin that couples to three neighbours with $J_{ij}=\pm 1$.
\begin{verbatim}
l0 = jzw0 ^ (spin ^ neighbour0.spin)
l1 = jzw1 ^ (spin ^ neighbour1.spin)
l2 = jzw2 ^ (spin ^ neighbour2.spin)
i1 = l0 ^ l1
i0 = i1 ^ l2
i1 = (l0 & l1) ^ (i1 & l2)
double u = rand(1)
if (u < p0) {
  spin = spin ^ (-1)
} else if (u < p1) {
  spin = spin ^ (i1 | i0)
} else {
  spin = spin ^ i1
}
\end{verbatim}
Here {\tt spin}, {\tt jzw0}, {\tt jzw1}, {\tt jzw2}, {\tt l0},
{\tt l1}, {\tt l2}, {\tt i0}, and {\tt i1}
are 64-bit words, {\tt spin} stores 64 spins
for 64 replicas, {\tt jzw} represents the coupling constant $J$ (all
bits of {\tt jzw} are set to zero if $J=1$ and all bits of {\tt jzw}
are set to one if $J=-1$). In the first three lines of the code, we
determine whether the interaction energy is positive or negative
for every pair of interacting spins. A bit of {\tt lj}
is set to one if the corresponding interaction energy is positive and
set to zero if the interaction energy is negative. In the next three lines
of the code, we calculate the index $i$ by bitwise summation of {\tt l0},
{\tt l1}, and {\tt l2}. In this simple example, we need only two words to
store the index. All possible indices $i$ for one replica, corresponding
energies $E_i$, energy gains $\Delta E_i$, and probabilities $p_{t,i}$
are listed in table~\ref{tab:param}.
\begin{table}
\centering
\begin{tabular}{c|c|c|c}
Index $i$ & Energy $E_i$ & Energy gain $\Delta E_i$ & Probability $p_{t,i}$ \\
\hline
0 & $-3|J|$ & $6 |J|$ & $\exp(-6\beta |J|)$ \\
1 & $-1|J|$ & $2 |J|$ & $\exp(-2\beta |J|)$ \\
2 & $1|J|$ & $-2 |J|$ & 1 \\
3 & $3|J|$ & $-6 |J|$ & 1 \\
\end{tabular}
\caption{Indices, energies, energy gains when a spin is flipped,
and probabilities for one spin that couples to three neighbours
with $|J|=1$.}
\label{tab:param}
\end{table}
In the seventh line, we draw a uniformly distributed random number.
In the next lines, we compare it to the probabilities $p_{t,i}$ and flip
spins. It is easy to deduce from the correspondence between indices
and probabilities which spins should be flipped. Namely, we find by simple
boolean logic the spins with indices $i \geq k$ such that
$p_{k-1,t} < u < p_{k,t}$ and flip these spins.
As the probabilities $p_{i,t}$ are used
often throughout the simulations, these are precomputed when the
algorithm is initialised.

\paragraph{Correlations}
This approach results in correlations between the replicas because
only one random number is used per update for all the $64$ replicas in
a word. For example, if at any point during the annealing two replicas
are in the same state, they will follow the same path, making one of
the replicas redundant. In the extreme case of fully correlated
replicas, all of them find the same state. Correlations can be
measured by computing the correlation ratio
$$ C = \frac{V}{W(1-W)} $$
over multiple repetitions of the annealing process, where
$W=(1/R)\sum_{i=1}^R w_i$ is the mean success rate (the probability of
finding the ground state), $R$ is the number of repetitions,
$V=(1/R)\sum_{i=1}^R (w_i - W)^2$ is the variance and
$w_i=(1/64)\sum_{j=1}^{64} w_{ij}$ is the mean success rate of the
$i$th repetition (each repetition has $S=64$ replicas). It can be shown
that $C$ is close to zero for uncorrelated replicas and $C=1$ for fully
correlated replicas.

\begin{figure}
\centering
\includegraphics[width=0.48\columnwidth]{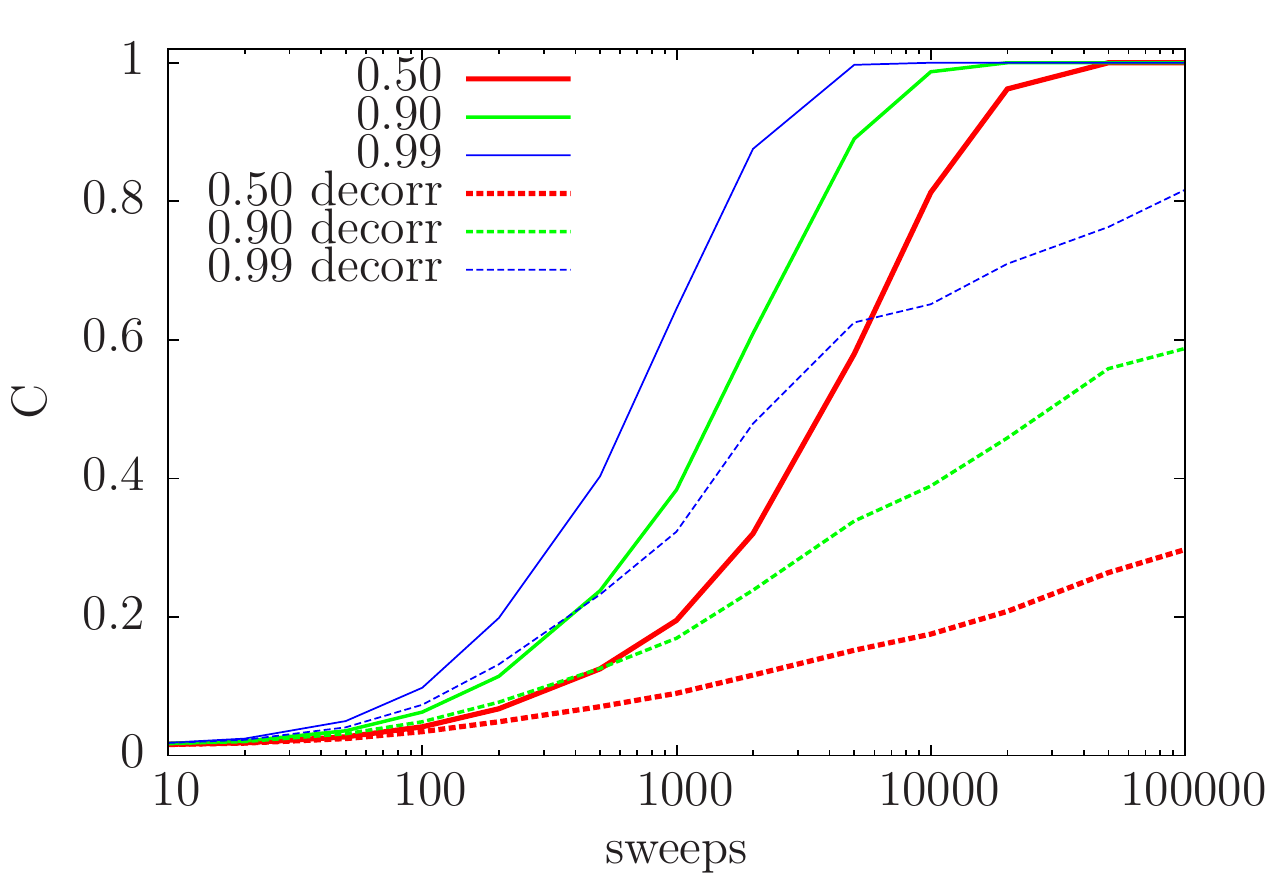}
\includegraphics[width=0.48\columnwidth]{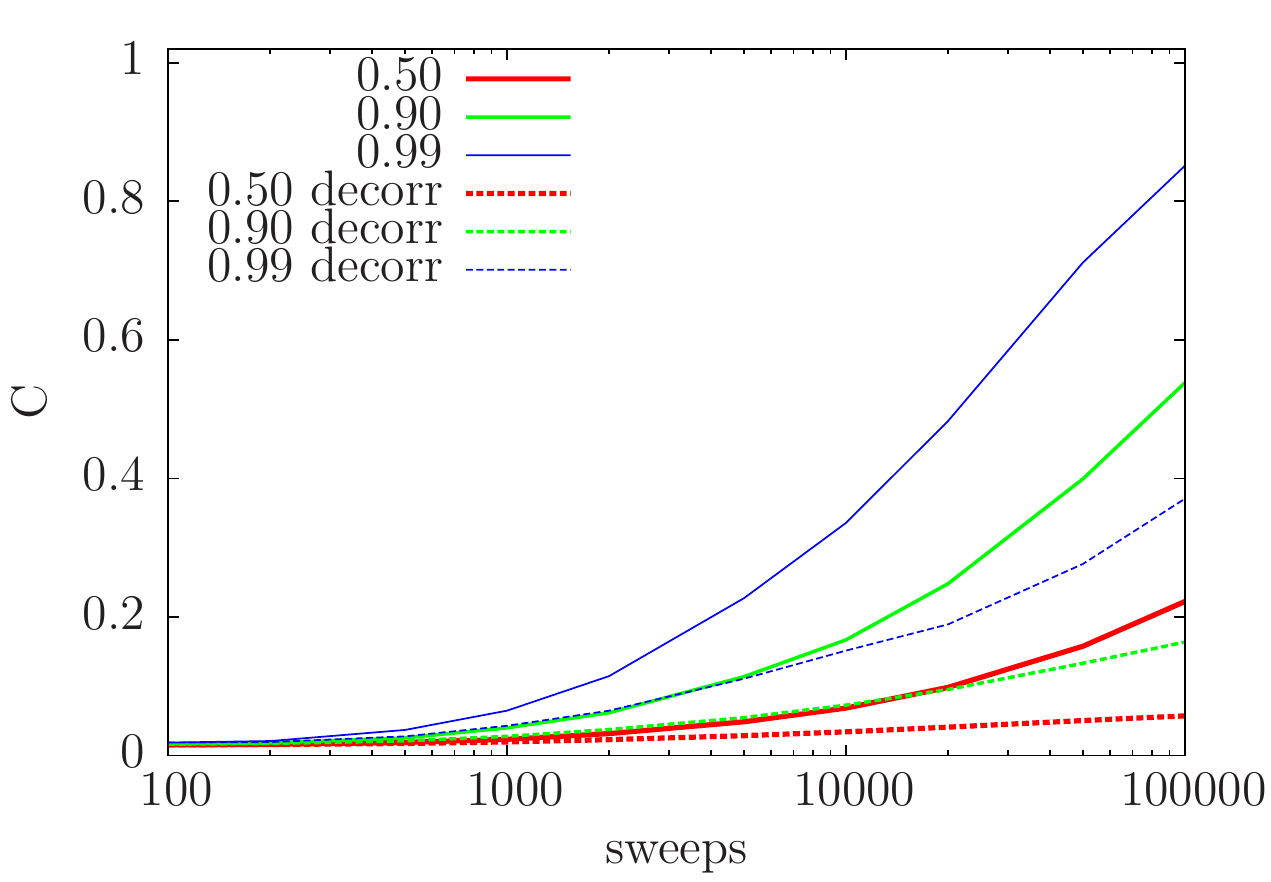}
\caption{The correlation ratio $C$ as a function of the number of
  sweeps for instances without fields on depleted chimera graphs with
  126 spins in the left panel and 503 spins in the right panel
  (see \ref{sec:chimera} for details).
  The data are presented for the 0.5, 0.9 and 0.99 quantiles.
  The definition of quantile is given at the beginning of
  Sec.~\ref{subsec:annealing:times}.
  The dotted and solid curves show the correlations
  introduced by the code with and without applying the decorrelation
  strategy respectively.}
\label{fig:correlations}
\end{figure}

In figure~\ref{fig:correlations}, we show the correlation ratio $C$ as
a function of the number of sweeps for instances on the chimera
graph. Correlations increase with the ratio of the number of sweeps to
the system size. However, they can be substantially reduced by not
flipping one random spin (bit) in each update. The random number that
is used to make an update can be reused. As can be seen, our
decorrelation strategy reduces the correlations significantly but not
fully. It should be emphasised that these correlations are usually
irrelevant because strong correlations appear only when the number of
sweeps is much larger than the optimal number of sweeps to find the
ground state with very high probability, as discussed below.

\subsubsection{Approach Two}
\label{sec:approach-two}
The second approach follows the ordinary algorithm where a spin is
picked, its local energy is computed and it is flipped with
probability $p$. However, instead of just flipping one spin, one
determines the individual energies of $64$ spins simultaneously and
computes whether the spins should be flipped from a set of
probabilities. The average case complexity of generating $Q$ $1$-bits
with probability $p$ in parallel is ${\cal
  O}(\log_2(Q)+2)$~\cite{springerlink:10.1007/BF01010404}. This way
$64$ bits are generated in on average $8$ iterations. Since we are
considering sparse graphs with a limited range, only a limited number
of flipping probabilities can be attained by a spin at each time step
and this makes the generalisation of the above algorithm to individual
flipping probabilities $p^{(1)},p^{(2)},\dots,p^{(Q)}$ straight
forward. While this algorithm is more than a factor of three slower
than the one presented in Sec.~\ref{sec:approach-one}, correlations
are here of the order of the pseudo random number generator.

\section{Optimising annealing strategies}
It is important to optimise both the slope and the length of the
annealing schedule.  

\subsection{Optimising the schedules}
We follow the ideas in Ref. \cite{wes}. Based on considerations of
keeping the average energy difference between two successive steps $k$
and $k+1$ below a threshold $\langle E_{k+1}\rangle - \langle E_k
\rangle \leq -\lambda\,\sigma_k$ it can be shown that $\beta_{k+1} =
\beta_{k} + \lambda \sigma^{-1}_k $ where $\sigma_k$ is the standard
deviation of the energy at step $k$ and $\lambda \leq 1$. We optimised
the schedule using this approach, but found that the performance was
only slightly better than a linear schedule. Much more important is
that the starting value of the temperature $T$ is around the same
order as the maximal energy required to flip any spin and that the end
value is low enough such that the state does not jump out of the final
minimum. For bi-modal couplings $J_{ij} = \pm1$, we found that the
inverse temperatures $\beta_s = 0.1 $ and $\beta_e = 3$ were good
initial and final values up to $512$ spin problems, and these values
were used for the benchmark runs.

\subsection{Optimising annealing times}
\label{subsec:annealing:times}

\begin{figure}
  \centering
  \includegraphics[width=0.5\columnwidth]{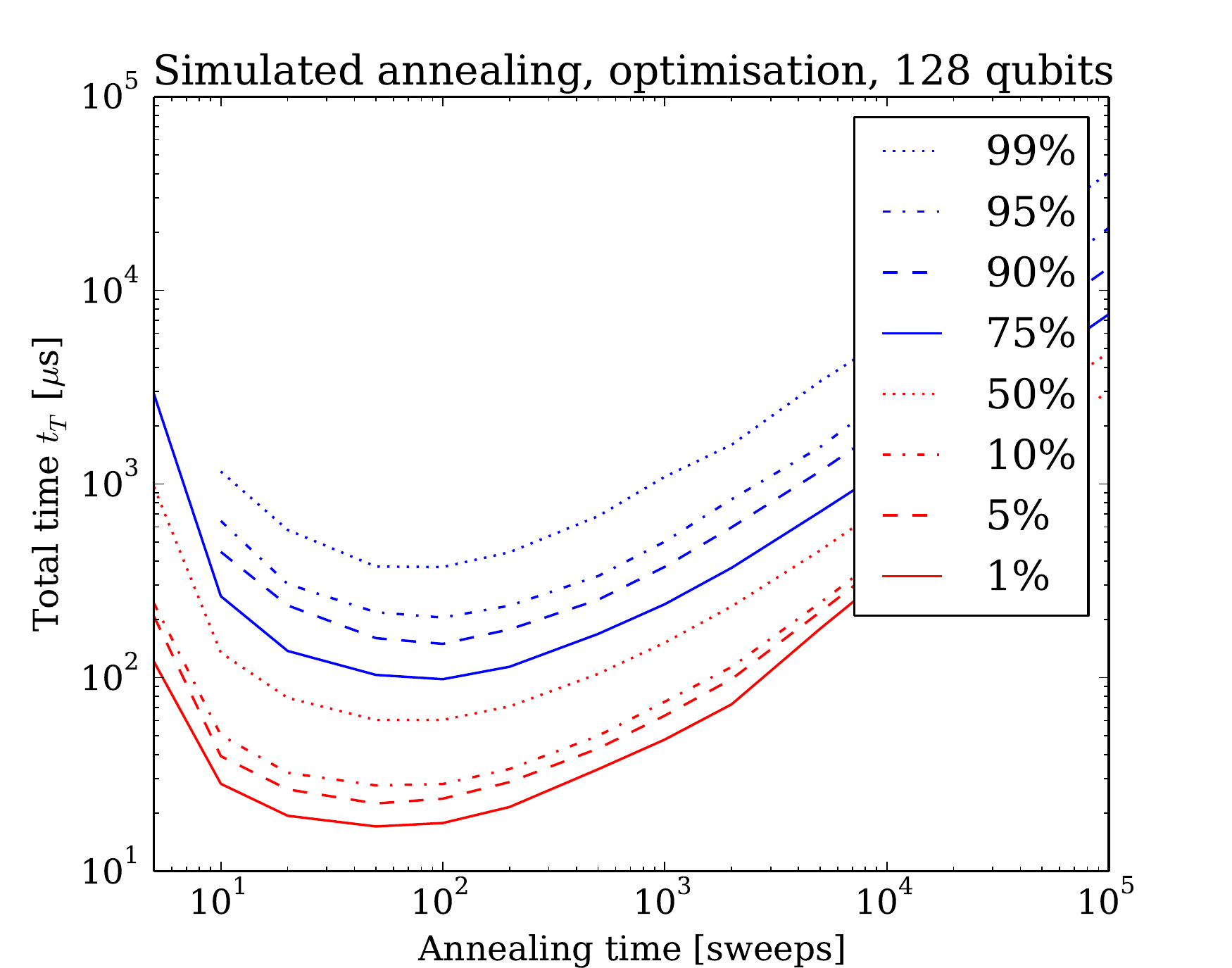}
  \caption{Quantile optimisation for $\pm J$ instances on the chimera graph.}
  \label{fig:optquant}
\end{figure}

As simulated annealing is a heuristic algorithm, one can strive
towards maximising the probability of finding the ground state by
either increasing the number of sweeps, increasing the number of
repetitions, or both at once. The optimal choice which achieves this
goal with minimal total computational effort depends on the the class
of problem at hand. Furthermore, for a set of such problems, how
should one run the code to lower the computational resources needed
for $5\%$ easiest ones?  For the $50\%$ easiest? Or for the
$99\%$-easiest? We refer to the various percentages as {\it
  quantiles}, and to address these questions, we consider an annealer
which is ran using $S$ sweeps with $R$ repetitions. If the probability
of finding the ground state for a single repetition is $w(S)$, then
the total number of repetitions needed to find the ground state with
$99\%$ is
\begin{equation}
  R = \left\lceil
    \frac{\log (0.01)}{\log(1 - w)} 
    \right\rceil ,
    \label{eq:repetitions}
\end{equation}
where we take the ceiling $\lceil\, \rceil$ as the number of repetitions of
annealing must be an integer. This gives a total annealing time of 
\begin{equation}
  \label{eq:tt}
  t_T = t_a \cdot R .
\end{equation}
Here $t_a$ is the annealing time for one repetition which is given by
$t_a = S \cdot N / f $ (when $S \cdot N$ is large), where $S$ is the
number of sweeps performed in the simulation, $N$ is the system size
and $f$ is the number of attempted spin updates per second. Since
$w(t_a)$ is a non-trivial function of $t_a$, the total time $t_T$ to
find the ground state with $99\%$ probability is a non-trivial
function of $t_a$, and therefore one needs to minimise $t_T$ as a
function of $t_a$ in order to find the optimal running parameters for
the algorithm. As an example here consider $1000$ random problems on
the chimera graph (\ref{sec:chimera}). We then plot $t_T(t_a)$ in
\fig{optquant}. It is evident that the code runs optimally when
ran between $t_a = 400 \cdot N /f$ and $t_a = 1000 \cdot N /f$ depending
on which quantile the problem belongs to.

We also consider the mean time to find the ground state \cite{McGeoch},
which is given by
\begin{equation}
  \label{eq:ts}
  t_M = t_a \left\lceil
    \frac{1}{w} 
    \right\rceil .
\end{equation}
That is we replaced $\log (0.01)/\log(1 - w)$ by $1/w$ in eq.~\ref{eq:tt}.
If $w$ is small then the mean time to find the ground state is a time
to find the ground state with $63\%$ probability.

\section{Simulated annealing codes}
\label{sec:codes}

We provide a number of simulated annealing codes. The codes are listed
in table~\ref{tab:codes}. Most of the multi-spin codes
are designed for lattices that have from one to six neighbours at each
site and with couplings that can be $\pm1$. The codes can be sped up by choosing the required
number of neighbours in the {\tt ms\_config.h} file. For instance, one
can leave just the {\tt \#define USE\_4\_NEIGHB} line and comment out the other
five lines for the square lattice.

We have also
developed multi-spin codes for other interaction ranges (by range-$n$
we mean integer couplings within a range $[-n, -n + 1, \ldots, -1, 1,
  \ldots, n - 1, n]$) and for lattices with larger (fixed) number of
neighbours. The programming of higher range codes is rather tedious, so
we use a code generator to generate these codes. We include  range-3 code for up to six neighbours. The other codes can be
obtained from the authors.

There are two sets of single-spin codes. The codes that end with
{\tt vdeg} use variable loop lengths over neighbours and designed for
any number of neighbours. The other codes use fixed loop lengths and
the maximum number of neighbours can be set in the {\tt ss\_config.h} file
(the default value is six). The latter set of codes can be faster in some
circumstances as discussed in Sec.~\ref{sec:fixed:loops}.

\begin{table}
\centering
\begin{tabular}{l|l}
Code & Description \\ \hline
{\tt an\_ms\_r1\_nf} & Multi-spin code for range-1 interactions without magnetic field (approach one). \\
{\tt an\_ms\_r1\_fi} & Multi-spin code for range-1 interactions with magnetic field (approach one). \\
{\tt an\_ms\_r3\_nf} & Multi-spin code for range-3 interactions without magnetic field (approach one). \\
{\tt an\_ms\_r1\_nf\_v0} & Multi-spin code for range-1 interactions without magnetic field (approach two). \\
{\tt an\_ss\_ge\_fi} & Single-spin code for general interactions with magnetic field (fixed number of neighbours). \\
{\tt an\_ss\_ge\_fi\_vdeg} & Single-spin code for general interactions with magnetic field (any number of neighbours). \\
{\tt an\_ss\_ge\_nf\_bp} & Single-spin code for general interactions on bipartite lattices without magnetic field \\ & (fixed number of neighbours). \\
{\tt an\_ss\_ge\_nf\_bp\_vdeg} & Single-spin code for general interactions on bipartite lattices without magnetic field \\ & (any number of neighbours). \\
{\tt an\_ss\_ge\_fi\_bp\_vdeg} & Single-spin code for general interactions on bipartite lattices with magnetic field \\ & (any number of neighbours). \\
{\tt an\_ss\_rn\_fi} & Single-spin code for range-$n$ interactions with magnetic field (fixed number of neighbours). \\
{\tt an\_ss\_rn\_fi\_vdeg} & Single-spin code for range-$n$ interactions with magnetic field (any number of neighbours).
\end{tabular}
\caption{A list of simulated annealing codes. These are also targets (executable names) for the Makefile. }
\label{tab:codes}
\end{table}

To find the ground state, annealing must be repeated many times as described
is Sec.~\ref{subsec:annealing:times}. The loop over repetitions can be
easily parallelized using OpenMP as the repetitions are independent of each
other. This can lead to a significant speedup (up to a factor of the number
of threads used) when the number of repetitions is large enough.
We provide both single-threaded and multi-threaded versions of the codes. 

\subsection{Building the codes}

The codes are all stand-alone and can be built using a \verb|C++11|
conforming compiler as follows: \verb|make target|, where
\verb|target| specifies the code to build. All available targets are
listed in table~\ref{tab:codes}. To build a multi-threaded version
append \verb|_omp| to \verb|target|.

\subsection{Running the codes}

\begin{table}
  \centering
  \begin{tabular}{l|l}
    Argument flag & Description \\\hline
    {\tt -l [instance]} & {\tt [instance]} specifies instance file. \\
    {\tt -s [sweeps]}   & {\tt [sweeps]} is number of sweeps. \\
    {\tt -r [reps]}     & {\tt [reps]} is number of repetitions. \\
    {\tt -b0 [beta0]}   & {\tt [beta0]} is initial inverse temperature. {\em Default value: {\tt 0.1}}. \\
    {\tt -b1 [beta1]}   & {\tt [beta1]} is final inverse temperature. {\em Default value: {\tt 3.0}}.\\
    {\tt -r0 [rep0]}    & {\tt [rep0]} is starting repetition. {\em Default value: 0}.\\ 
    {\tt -v} & if {\tt -v} is set, timing and some other info is printed. {\em Default value: not set}.\\
    {\tt -g} & if {\tt -g} is set, only the lowest energy solution is printed. {\em Default value: not set}.\\
    {\tt -sched [schedule]} & {\tt [schedule]} specifies a schedule. It can either be {\tt lin}, \\
                     & {\tt exp} or be a text file on the system which contains an \\
                     & inverse temperature on every line. {\em Default value: {\tt lin}}.\\
    {\tt -t [threads]}  & {\tt [threads]} is the number of threads to run in parallel. {\em Default value: {\tt OMP\_NUM\_THREADS}}.
  \end{tabular}
  \caption{Command-line arguments of the provided algorithms.}
  \label{tab:descrun}
\end{table}

Either of the previously described algorithms can be ran by using the
appropriate executable, see table~\ref{tab:codes}.  Every algorithm
follows the same command-line interface which allows one to specify
the lattice using {\tt -l}, the schedule {\tt -sched}, the number of
sweeps {\tt -s} and the number of repetitions {\tt -r}. Because
repetitions are independent of each other, the codes can be trivially
parallelised. The number of threads to run in parallel can be
specified by {\tt -t}. In cases where one uses one of the
pre-programmed schedules, {\tt lin} or {\tt exp}, one can also specify
the initial inverse temperature $\beta_0$ and the final inverse
temperature $\beta_1$. In table~\ref{tab:descrun} we summarise the
full set of command-line arguments.  A custom schedule can be loaded
by using {\tt -sched} followed by the name of a text file.

The input lattice files are plain text files with following structure:
First line is the name of the lattice, and following $L+M$ lines
contain $L$ couplings and $M$ local fields (not ordered). Each line
contains three values $i$, $j$ and $c$. If $i=j$ the line specifies a
local field on site $i$ of size $h_i=c$. Otherwise, the line denotes a
coupling between spin $i$ and $j$ of value $J_{ij}=c$.

The output contains four columns: the energy, the number of times
this energy is found, the success rate (the second column divided by
the total number of repetitions), and the instance file name.

Sample input and output data are located in the {\tt example}
sub-directory of the source code distribution.

\section{Benchmarking}
\label{sec:benchmarking}

We benchmark our codes on a 8-core Intel Sandy Bridge Xeon E5-2670
processor with hyper-threading enabled for Ising spin glass instances
on the chimera graph. The codes are compiled by the GNU C++ compiler,
version 4.7.2. To benchmark the codes, $1000$ range-$1$ random
$503$-spin instances without fields are generated and annealing times
are optimized as described in Sec.~\ref{subsec:annealing:times}.
The ground state energies of all instances were verified by exact
solvers~\cite{speedup}.
In multi-spin versions of the code, we run $64$ repetitions
simultaneously. Thus we have to round the number of repetitions to the
nearest largest values that are multiples of $64$.

Generically, the multi-threaded version of the multi-spin code shows
the best performance when the number of repetitions that is needed to
find the ground state is large enough.

Times to find the solution with $99\%$ probability, given by
eq.~\ref{eq:tt}, for the $99\%$ quantile, see
Sec.~\ref{subsec:annealing:times}, are reported in
tables~\ref{tab:bench:chimera:a}, \ref{tab:bench:chimera:b} and
\ref{tab:bench:chimera:c}.  Table~\ref{tab:bench:chimera:b} shows that
we can reach $50$ spin flips per nanosecond on a single Intel
processor using our fast multi-spin and multi-threaded code. For the
$99\%$ quantile on a $503$-chimera graph, the time to find the
solution is less than $153$ ms for range-1 instances without random
fields and less than 27ms for instances with random fields.

Mean times to find the solution, given by eq.~\ref{eq:ts}, for the
$99\%$ quantile are reported in tables~\ref{tab:bench:chimera:d},
\ref{tab:bench:chimera:e} and \ref{tab:bench:chimera:f}. The mean time
to find the solution is $8.2$ ms and less than $40$ ms for instances
with and without fields for the fastest code.  The time for instances
with local random fields can be compared to the mean times to the
solution reported for similar benchmarks on smaller problems in
reference~\cite{McGeoch}.

\begin{table}
\centering
\begin{tabular}{l|r|r|r|r|r}
  Code & Sweeps & Repetitions & Init time in ms & Run time in ms & Spin flips per ns \\ \hline
     {\tt an\_ms\_r1\_nf} & 2000 & 7872 &   0.6 &  1190 & 6.65 \\
{\tt  an\_ms\_r1\_nf\_v0} & 2000 & 7872 &   2.9 &  4172 & 1.90 \\
	{\tt  an\_ss\_ge\_fi} & 2000 & 7858 &  69.0 & 26206 & 0.30 \\
 {\tt an\_ss\_ge\_nf\_bp} & 1000 & 2301 &  19.4 &  6708 & 0.09 \\
     {\tt an\_ss\_rn\_fi} & 2000 & 7858 &   0.7 & 25590 & 0.31
\end{tabular}
\caption{Time $t_T$ to find the solution with probability 0.99 for the 99\% quantile for range-$1$ random instances without fields on the chimera graph of size 503 using 1 thread.}
\label{tab:bench:chimera:a}
\end{table}

\begin{table}
\centering
\begin{tabular}{l|r|r|r|r|r}
  Code & Sweeps & Repetitions & Init time in ms & Run time in ms & Spin flips per ns \\ \hline
     {\tt an\_ms\_r1\_nf} & 2000 & 7872 &  0.7 & 151.7 & 52.2 \\
{\tt  an\_ms\_r1\_nf\_v0} & 2000 & 7872 &  139 & 491.1 & 16.1 \\
    {\tt  an\_ss\_ge\_fi} & 2000 & 7858 &  222 &  3261 & 2.42 \\
 {\tt an\_ss\_ge\_nf\_bp} & 1000 & 2301 & 58.0 & 876.6 & 0.66 \\
     {\tt an\_ss\_rn\_fi} & 2000 & 7858 &  0.8 &  3177 & 2.49
\end{tabular}
\caption{Time $t_T$ to find the solution with probability 0.99 for the 99\% quantile for range-$1$ random instances without fields on the chimera graph of size 503 using 16 threads.}
\label{tab:bench:chimera:b}
\end{table}

\begin{table}
\centering
\begin{tabular}{l|r|r|r|r|r}
  Code & Sweeps & Repetitions & Init time in ms & Run time in ms & Spin flips per ns \\ \hline
     {\tt an\_ms\_r1\_fi} & 1000 & 1792 &  0.8 &  25.4 & 35.5 \\
    {\tt  an\_ss\_ge\_fi} & 1000 & 1732 &  112 & 340.0 & 2.56 \\
     {\tt an\_ss\_rn\_fi} & 1000 & 1732 &  0.8 & 335.6 & 2.59
\end{tabular}
\caption{Time $t_T$ to find the solution with probability 0.99 for the 99\% quantile for range-$1$ random instances with fields on the chimera graph of size 503 using 16 threads.}
\label{tab:bench:chimera:c}
\end{table}

\begin{table}
\centering
\begin{tabular}{l|r|r|r|r|r}
  Code & Sweeps & Repetitions & Init time in ms & Run time in ms & Spin flips per ns \\ \hline
     {\tt an\_ms\_r1\_nf} & 2000 & 1728 &  0.6 & 261.2 & 6.65 \\
{\tt  an\_ms\_r1\_nf\_v0} & 2000 & 1728 &  2.9 & 916.0 & 1.90 \\
	{\tt  an\_ss\_ge\_fi} & 2000 & 1667 & 68.7 &  5556 & 0.30 \\
 {\tt an\_ss\_ge\_nf\_bp} &  500 & 1000 & 10.8 &  1469 & 0.09 \\
     {\tt an\_ss\_rn\_fi} & 2000 & 1667 &  0.7 &  5431 & 0.31
\end{tabular}
\caption{Mean time $t_M$ to the solution for the 99\% quantile for range-$1$ random instances without fields on the chimera graph of size 503 using 1 thread.}
\label{tab:bench:chimera:d}
\end{table}

\begin{table}
\centering
\begin{tabular}{l|r|r|r|r|r}
  Code & Sweeps & Repetitions & Init time in ms & Run time in ms & Spin flips per ns \\ \hline
     {\tt an\_ms\_r1\_nf} & 2000 & 1728 &  0.7 &  38.4 & 45.2 \\
{\tt  an\_ms\_r1\_nf\_v0} & 2000 & 1728 &  139 & 122.9 & 14.1 \\
    {\tt  an\_ss\_ge\_fi} & 2000 & 1667 &  224 & 696.8 & 2.40 \\
 {\tt an\_ss\_ge\_nf\_bp} &  500 & 1000 & 30.0 & 189.8 & 0.66 \\
     {\tt an\_ss\_rn\_fi} & 2000 & 1667 &  0.8 & 678.9 & 2.47
\end{tabular}
\caption{Mean time $t_M$ to the solution for the 99\% quantile for
  range-$1$ random instances without fields on the chimera graph of
  size 503 using 16 threads.}
\label{tab:bench:chimera:e}
\end{table}

\begin{table}
\centering
\begin{tabular}{l|r|r|r|r|r}
  Code & Sweeps & Repetitions & Init time in ms & Run time in ms & Spin flips per ns \\ \hline
     {\tt an\_ms\_r1\_fi} & 2000 &  384 &  0.7 &  7.5 & 25.7 \\
    {\tt  an\_ss\_ge\_fi} & 2000 &  377 &  111 & 75.4 & 2.51 \\
     {\tt an\_ss\_rn\_fi} & 2000 &  377 &  0.8 & 74.5 & 2.54
\end{tabular}
\caption{Mean time $t_M$ to the solution for the 99\% quantile for
  range-$1$ random instances with fields on the chimera graph of
  size 503 using 16 threads.}
\label{tab:bench:chimera:f}
\end{table}

\section{Acknowledgements}

This work was supported by the Swiss National Competence Center in
Research on Quantum Science and Technology NCCR-QSIT and by the
European Research Council through the ERC grant SIMCOFE. We
acknowledge discussions with I. Pizorn, L. Gamper and
C. McGeoch. M.T. acknowledges hospitality of the Aspen Center for
Physics, supported by NSF grant PHY-1066293.

\appendix
\section{The chimera graph}
\label{sec:chimera}
\begin{figure}
  \centering
    \includegraphics[width=0.3\columnwidth]{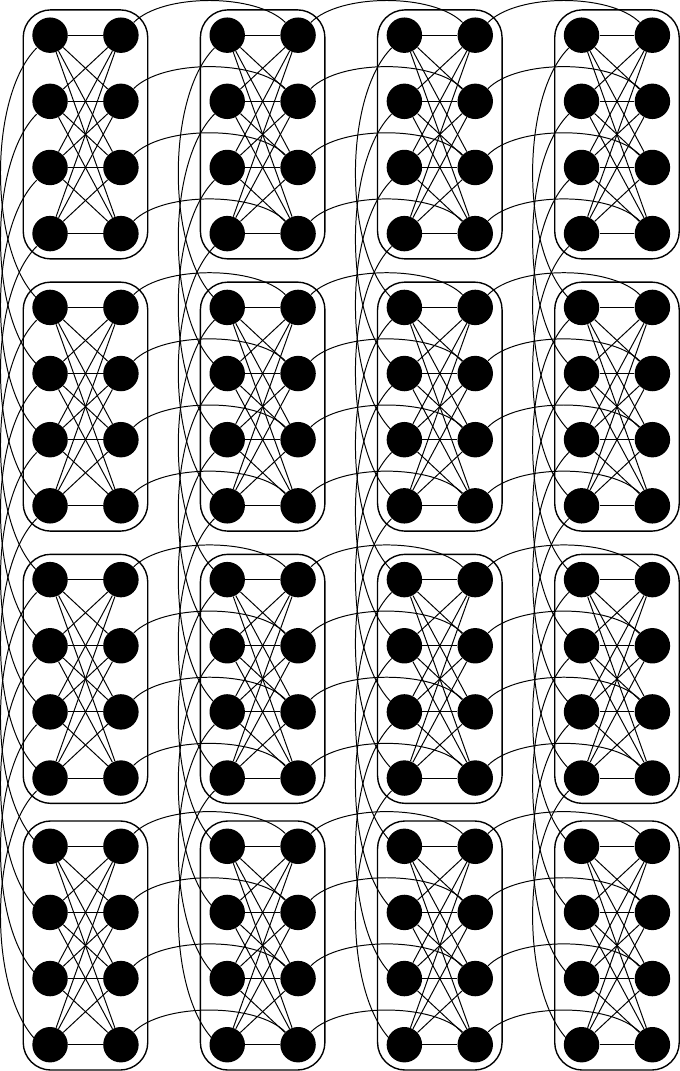}
  \caption{$4\times 4$ $K_{4,4}$ chimera graph having a total of 128
    vertices.}
  \label{fig:chimera}
\end{figure}

Chimera graphs are lattices with bipartite fully connected unit cells
consisting of $2c$ vertices (denoted by $K_{c,c}$) distributed on a $L
\times L$ grid - see figure~\ref{fig:chimera}. D-Wave Two implements a
transverse field Ising model on an $8\times 8$ $K_{4,4}$ graph
where every node represents a spin and every edge represents a
coupling. We have used this graph for benchmarking our codes. In order
to allow direct comparisons to an actual D-Wave device we use chimera
graphs with missing vertices (due to fabrication issues) as on the
D-Wave Two device located at the University of Southern
California. Example input files using the specific 126 and 503 spin
graphs used here are included in the software package.

\bibliographystyle{elsarticle-num}
\bibliography{paper}

\end{document}